\documentclass[%
 reprint,
superscriptaddress,
 amsmath,amssymb,
 aps,
]{revtex4-2}
\usepackage{hyperref}
\usepackage{graphicx}% Include figure files
\usepackage{dcolumn}% Align table columns on decimal point
\usepackage{bm}% bold math
\usepackage{amsmath}  % improve math presentation
\usepackage[margin=1in,letterpaper]{geometry} % decreases margins
\usepackage{physics}
\usepackage{import}
\usepackage[caption=false]{subfig}
\hypersetup{
	colorlinks=true,       % false: boxed links; true: colored links
	linkcolor=blue,        % color of internal links
	citecolor=blue,        % color of links to bibliography
	filecolor=magenta,     % color of file links
	urlcolor=blue         
}

\begin{document}

\preprint{APS/123-QED}

\title{Constructive Quantum Interference in a Photo-Chemical Reaction of $^{87}$Rb Bose Einstein Condensate}% Force line breaks with \\
%\thanks{A footnote to the article title}%
\author{Sumit Suresh Kale}
\affiliation{Department of Chemistry, Purdue University, West Lafayette, IN 47906, USA}
\author{Yong P. Chen}
\affiliation{Department of Physics and Astronomy, Purdue University, West Lafayette, USA}
\affiliation{Purdue Quantum Science and Engineering Institute, Purdue University, West Lafayette, USA}
\author{Sabre Kais}
\email{Corresponding author: kais@purdue.edu}
\affiliation{Department of Chemistry, Purdue University, West Lafayette, IN 47906, USA}
\affiliation{Department of Physics and Astronomy, Purdue University, West Lafayette, USA}
\affiliation{Purdue Quantum Science and Engineering Institute, Purdue University, West Lafayette, USA}
%\author{$\text{Sumit Suresh Kale}^_{*}$, $\text{Yong P. Chen}^_{\dagger,\ddagger}$, $\text{Sabre Kais}^_{*,\dagger,\ddagger}$}
% \affiliation{$^_{*}$Department of Chemistry, Purdue University, West Lafayette, IN 47906, U.S.A., $^_{\dagger}$Department of Physics and Astronomy, Purdue University, West Lafayette, U.S.A., $^_{\ddagger}$Purdue Quantum Science and Engineering Institute, Purdue University, West Lafayette, USA }%Lines break automatically or can be forced with \\

\date{\today}% It is always \today, today,
             %  but any date may be explicitly specified

\begin{abstract}
Interferences emerge when multiple pathways coexist together leading towards the same result. A recent study by Blasing and coworkers [PRL 121(7):073202] showed that in a photo-association reaction of Raman dressed spin-orbit coupled $^{87}$Rb Bose Einstein Condensate when the reactant spin state is prepared in a coherent superposition of multiple bare spin states it leads to a destructive interference between reaction pathways. Here we report a theoretical study for a reaction scheme that leads to constructive quantum interference. This is achieved by changing the reactive scattering channel in the reaction. As the origin of coherent control comes from the spin part of the wavefunction it is sufficient to use radio frequency coupling to achieve the superposition state. Our results show that interferences can be used as a resource for the coherent control of photo-chemical reactions. The approach is general and can be employed to study a wide spectra of chemical reactions in the ultracold regime.
\end{abstract}

%\keywords{Suggested keywords}%Use showkeys class option if keyword
                              %display desired
\maketitle

%\tableofcontents

Observation of quantum interference is a common phenomenon in the physics of microscopic particles such as atoms and electrons, but when it comes to chemical reactions it is quite rare. In recent years several experiments \cite{jambrina2015quantum,dai2003interference,sneha2016multiple}  have shown that if there exists multiple pathways for a chemical reaction then these pathways may interfere with each other producing interference patterns. Similar experiment has been carried out by Blasing et. al. \cite{blasing2018observation} which tried to answer the question what happens to the reaction when reactants are prepared in a superposition?

Photoassociation (PA) \cite{jones2006ultracold} is a light aided chemical process where two atoms absorb a photon producing a bounded excited molecule while scattering. When we operate PA in ultracold regime it involves only a small number of scattering channels. Experimentally, a Bose Einstein Condensate (BEC) of $^{87}$Rb can be prepared in $f=1$ hyperfine state via optical evaporation \cite{olson2013optimizing}. The magnetic field tuning during the optical evaporation results in a BEC with bare $m_{f}=-1,0,+1$ spin states or a statistical mixture of all three. In our calculations we consider a BEC of $^{87}$Rb in $f=1$ and $m_{f}=0$  bare spin state. 
\section*{Using Raman Coupling to Achieve Superposition}
A superposition state can be experimentally achieved by applying two counter-propagating Raman lasers adiabatically which drives transitions between these atomic Zeeman levels \cite{socreview,socreview2}. As a result, the Rb atom makes a transition from $m_f$ to an $m_f-1$ hyperfine Zeeman state by absorbing and emitting a photon. This process induces, in addition, a change in the momentum of the atom by $2k_r$, where $k_r$ is the photon recoil momentum \cite{lin2009bose,lin2011spin}. Thus, the atoms are dressed into a superposition of hyperfine spin and mechanical momenta. In our previous work \cite{kale2020spin} we analyzed the spin orbit coupling in the BECs that realize a pair of qutrits. The Hamiltonian \cite{lin2011spin,lin2009bose} that describes such a spin($m_{f}$)- momentum ($K$) coupling can be written in the coupled basis $\ket{m_{f},K}=\{\ket{-1,q+2K_{r}},\ket{0,q},\ket{+1,q-2k_{r}}\}$ as:
\begin{equation}
\resizebox{.47 \textwidth}{!} 
{$H_0=\begin{pmatrix}
    \frac{\hbar^2}{2m}(q+2k_r)^2-\delta(B) && \frac{\Omega_{r}}{2} && 0 \\
    \frac{\Omega_{r}}{2} && \frac{\hbar^2}{2m}q^2-\epsilon(B) && \frac{\Omega_{r}}{2} \\
    0 && \frac{\Omega_{r}}{2} && \frac{\hbar^2}{2m}(q-2k_r)^2+\delta(B)
    \end{pmatrix}$}
\label{eq1}
\end{equation}
 Here $m$ is the mass of $^{87}Rb$, $q$ the quasi-momentum (usually at the minimum of the BEC's lowest energy band), $\Omega_{r}$ the strength of the Raman coupling (which determines the Rabi frequency for the Raman transition between two hyperfine $m_f$ states), $\delta(B)$ the detuning of the Raman laser, $\epsilon(B)=0.65E_{r}$ is the quadratic Zeeman shift ( at $|\vec{B}_{Bias}|\approx 5 G$) , $E_{r} = \frac{\hbar^{2}k_{r}^{2}}{2m}$ is the recoil energy and $B$ the strength of the external magnetic field.  In deriving Hamiltonian \eqref{eq1}, we made use of the rotating-wave approximation. After applying the Raman lasers the population is transferred to $m_f=1$ and $m_f=-1$ from initially created $m_f=0$ state due to which the BEC eventually ends up in the ground state of Hamiltonian \eqref{eq1}, described by: 
 \begin{equation}
    |\psi_0\rangle=C_{-1}|q+2k_r,-1\rangle+C_0|q,0\rangle+C_1|q-2k_r,1\rangle
    \label{eq2}
\end{equation}
Here, $C_{\pm1}$ and $C_{0}$ are coefficients of the superposition ground state as a result of coupling. The laser that drives a spin sensitive photo-association transition is then applied in the experiment\cite{blasing2018observation}, selectively photo-associating only those colliding atoms (denoted as $a$ and $b$ here) whose total angular momentum  $\ket{F=f_{a}+f_{b},m_{f}=m_{f,a}+m_{f,b}}$ = $\ket{0,0}$. Using the single particle basis, $\ket{f_{a},m_{f,a}}\ket{f_{b},m_{f,b}}$, $\ket{F=0,m_{f}=0} = (\ket{1,-1}\ket{1,+1}-\ket{1,0}\ket{1,0}+\ket{1,+1}\ket{1,-1})/\sqrt{3}$. After considering the indistinguishable nature of bosons we see that there are two pathways for this transition. Bosons with $m_{f}=\mp1$ and $m_{f}=\pm1$ combines together to give a molecule in $m_{f}=0$  and similarly two individual Bosons in $m_{f}=0$ does the same job. So the PA reaction happens through two pathways simultaneously. Both the reaction pathways contribute towards the total reaction rate with opposite signs due to opposite Clebsch-Gordon (CG) coefficients ($\pm 1/\sqrt{3}$ for $\ket{F=0,m_{f}=0}$) and the contribution also depends on the coefficients of the superpositioned states from Eq.\ref{eq2}. The rate of PA reaction $k_{PA} \propto |\bra{\psi_{mol}}\vec{d}.\vec{E}\ket{\psi_{scat}}|^{2}$, where the proportionality factor is independent of spin \cite{theis2004tuning}, $\psi_{mol}$ and $\psi_{scat}$ are the total molecular and scattering wavefunctions. $\vec{E}$ and $\vec{d}$  corresponds to the electric field of the PA laser and the dipole operator. Refs \cite{blasing2018observation,mckenzie2002photoassociation, blasing2}  give an in depth derivation of reaction rates calculation. As derived in Ref.\cite{blasing2} for the raman dressed atoms in the $\ket{F=0,m_{f}=0}$ scattering channel the ratio of reaction rates between atoms in superposition ($k_{sup}$) and bare spin ($k_{0,0}$) states is:
       \begin{equation}
    \frac{k_{sup}}{k_{0,0}} = |C_{0}^{2}|^{2} +4|C_{-1}C_{1}|^{2}-4 Re[C_{0}^{2}C_{-1}^{*}C_{1}^{*}]
    \label{des}
\end{equation}
The last term becomes negative because of the opposite sign of CG coefficients. $\Omega_{r}=0$ and $\delta=0$ can be visualized as not applying the raman coupling beams and thus we no longer have the superposition in the reactant. This corresponds to all the population of BEC in $\ket{f=1,m_{f}=0}$ hyperfine state ($C_{0} = 1, C_{\pm1}=0$ in Eq. \ref{des}) and thus   $k_{sup}/k_{0,0} \longrightarrow 1$. At large values of raman coupling and zero detuning half of the population is transferred equally to $\ket{f=1,m_{f}=1}$ and $\ket{f=1,m_{f}=-1}$ from $\ket{f=1,m_{f}=0}$ hyperfine spin state, which results in the convergence of the coefficients of the ground state  $C_{\pm{1}} \longrightarrow 1/2$ and $C_{0} \longrightarrow 1/\sqrt{2}$ and thus the reaction rate ratio (Eq. \ref{des})    $k_{sup}/k_{0,0} \longrightarrow 0$  (destructive interference).

Now consider what happens if we change the reaction scheme and use a PA reaction which selectively photo-associates only those colliding atoms (denoted as $a$ and $b$ here) whose total angular momentum  $\ket{F=f_{a}+f_{b},m_{f}=m_{f,a}+m_{f,b}}$ = $\ket{2,0}$. Using the single particle basis, $\ket{f_{a},m_{f,a}}\ket{f_{b},m_{f,b}}$, $\ket{F=2,m_{f}=0} = (\ket{1,-1}\ket{1,+1}+2\ket{1,0}\ket{1,0}+\ket{1,+1}\ket{1,-1})/\sqrt{6}$. The detailed theoretical discussion of the reaction rate ratio calculation for this scheme is available in the supplementary material. The ratio of reaction rates for this reaction scheme is:  

\begin{equation}
\frac{k_{sup}}{k_{0,0}} = \left|C_{0}^{2} \right|^{2}+\left|C_{1}C_{-1} \right|^{2} + 2\,Re[C_{0}^{2}C_{1}^{*}C_{-1}^{*}]
\label{cons}
\end{equation}
Here we see that due to same signs of the CG coefficients for the scattering channel $\ket{F=2,m_{F}=0}$ the last term comes out to be positive and this corresponds to constructive interference.  

\begin{figure}[h]
\includegraphics[width=1.\linewidth]{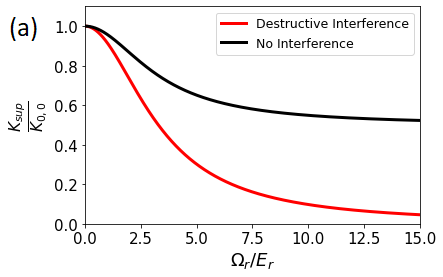}
 \quad
  \includegraphics[width=1.\linewidth]{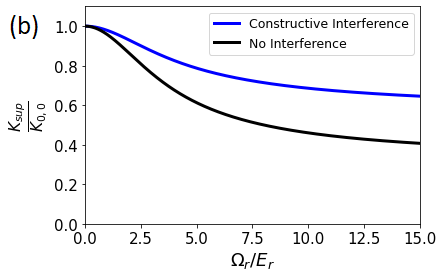}
 \caption{\label{Omega}%
  Photo-association rates ratio $k_{sup}/k_{0,0}$ of BEC, as a function of raman coupling $\Omega_{r}/E_{r}$ at detuning $\delta = 0\,E_{r}$, Fig.1a shows the curve for the channel $\ket{F=0,m_{f}=0}$. The black(red) curve corresponds to theoretical prediction without(with) the destructive interference term in the Eq. \ref{des}. The curve for the channel $\ket{F=2,m_{f}=0}$ is shown in Fig.1b. The black(blue) curve corresponds to theoretical prediction without(with) the constructive interference term in the Eq. \ref{cons}.
 }%
\end{figure}
 Fig.\ref{Omega}a shows the normalized photo-association rates ratio $k_{sup}/k_{0,0}$ of BEC for the \linebreak $\ket{F=0,m_{f}=0}$ channel, as a function of raman coupling $\Omega_{r}/E_{r}$ which ranges from 0 to 15 at detuning $\delta = 0\,E_{r}$. We see that when $\Omega_{r}/E_{r} = 0$ which is equivalent to no raman beam being applied and as a result we don't have any superposition in the reactant state, $C_{0}=1$ and $C_{\pm1}=0$ in Eq. \ref{des}, thus $k_{sup}/k_{0,0}\longrightarrow 1$. Superposition states induced by a large Raman coupling and zero Raman detuning, nearly complete suppression of the photoassociation rate is observed (red curve) which is interpreted as destructive interference. This result is consistent with the experiment \cite{blasing2018observation} where they observed a complete suppression of PA rection although the PA laser remained on. Fig. \ref{Omega}b  corresponds to $\ket{F=2,m_{f}=0}$ scattering channel. Where we see that the reaction rate  ratio for the case when we consider interference (blue curve) is always higher than the case without the interference term, which we interpret as constructive interference.

\begin{figure}
\includegraphics[width=1.\linewidth]{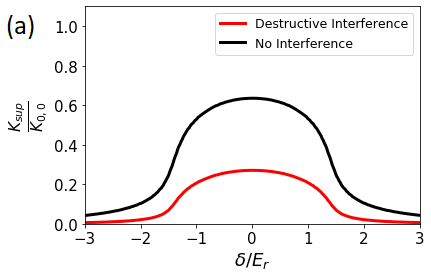}
 \quad
  \includegraphics[width=1.\linewidth]{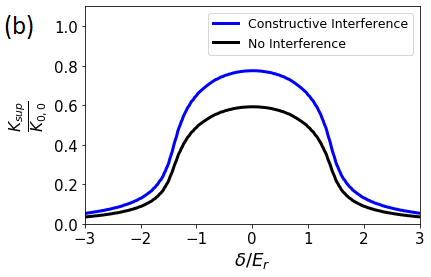}
 \caption{\label{Delta}%
  Photo-association rates ratio $k_{sup}/k_{0,0}$ as a function of $\delta/E_{r}$ at $\Omega_{r} = 5.4 E_{r}$, Fig.2a shows the curve for the channel $\ket{F=0,m_{f}=0}$. The black(red) curve corresponds to theoretical prediction without(with) the destructive interference term in the Eq. \ref{des} The curve for the channel $\ket{F=2,m_{f}=0}$ is shown in Fig.2b. The black(blue) curve corresponds to theoretical prediction without(with) the constructive interference term in the Eq. \ref{cons}.
 }%
\end{figure}

 We next study the effect of detuning $\delta/E_{r}$ on the PA rate. Since the BEC is prepared at the band minima, first the dressed band structure was calculated for $\Omega_{r} = 5.4 E_{r}$ and different values of $\delta/E_{r}$ and the quasimomentum values were obtained (corresponding to the minimum energy for a particular $\delta$ value). These values were used in Hamiltonian \ref{eq1} to obtain the superposition coefficients and then the reaction rate ratios were calculated. Fig. \ref{Delta} shows the normalized photo-association rates ratio $k_{sup}/k_{0,0}$ at different values of detuning $\delta/E_{r}$ ranging from -3 to 3 and at raman coupling $\Omega_{r} = 5.4 E_{r}$. Sub-figure \ref{Delta}a shows the results corresponding to the $\ket{F=0,m_{f}=0}$ channel where we see that the reaction rate ratio is always lower when we consider the interference (red curve) denoting destructive interference. The result is consistent with experimental findings by \cite{blasing2018observation}. In sub-figure\ref{Delta}b our result predict that the reaction rate ratio should be always higher in the case when we consider interference (blue) as compared to the no-interference case (black curve) which denotes constructive interference. Additionally it's worthwhile to note that the difference is highest between these two cases when the raman beam is resonant (the detuning $\delta$ is $0\,E_{r}$). It happens because when the raman beam is resonant it results in a better superposition and as we increase the detuning $\approx\pm2\,E_{r}$ the majority of population is transfered in either $m_{f}=\,\mp1$ which suggests that $C_{0}=0$ and one of $C_{\pm}=0$ which makes $k_{sup}/k_{0,0}\longrightarrow 0$  (Eq. \ref{des},Eq. \ref{cons}).

\section*{Using Radio Frequency to achieve Superposition}

\begin{figure}
\includegraphics[width=.7\linewidth]{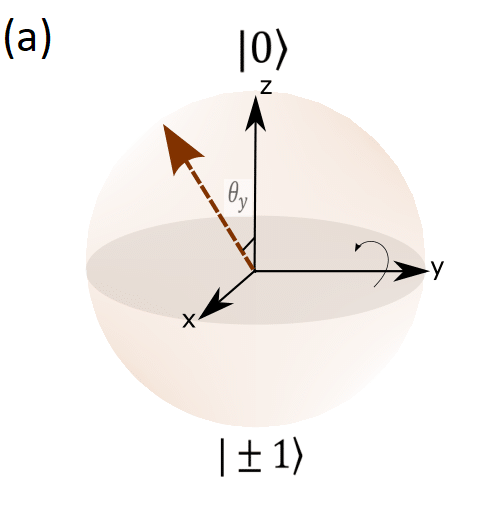}
 \quad
  \includegraphics[width=1.\linewidth]{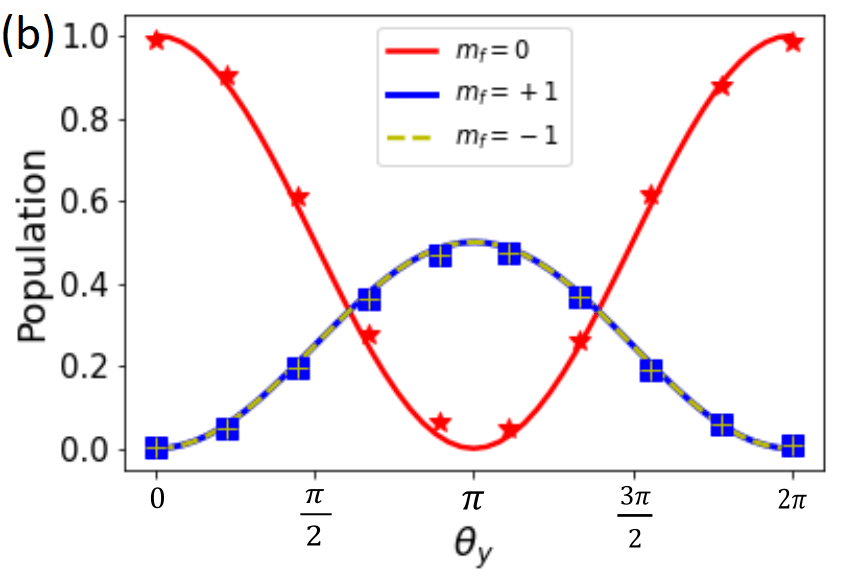}
 \caption{\label{fig:rf}%
  Achieving superposition using RF pulse, Fig.\ref{fig:rf}a shows the schematic representation of the rotation $\theta_{y}$ along Y axis on the Bloch sphere. Fig.\ref{fig:rf}b shows the redistribution of population as a function of $\theta_{y}$.
The red curve corresponds to state vector simulation of $m_{f}=0$ population and the asterisk data points on it corresponds to data obtained from the IBM quantum device. Similarly, the Blue(Yellow dash) cure correspond to $m_{f}=+1$(-1) and the square(plus) data points on it correspond to data obtained from IBM quantum device (IBMQ Lima).
 }%
\end{figure}

Since PA control only comes from the spin part of the superposition wavefunction, the momentum part created by the Raman beam is a distraction for underlying Physics. It is sufficient to use radio frequency to couple different $m_{f}$ spin states which we model below. The three-level hyperfine spin states can be schematically represented by a pair of bloch spheres \cite{kondakci2020interferometric}. Where one pole corresponds to $m_{f}=0$ and another pole corresponds to $m_{f}=\pm1$. Initially created BEC of $^{87}$Rb in $f=1$ and $m_{f}=0$  bare spin state is now coupled to the $m_{f}=\pm1$ states with a RF (Radio Frequency) field. By controlling the time for which the RF pulse is applied we can introduce a rotation along $Y$($\theta_{y}$) as shown in Fig. \ref{fig:rf}a. As a result of rotation, the population transfer takes place. We simulated the $\theta_{y}$ rotation via state vector simulator in Qiskit \cite{Qiskit} and confirmed the results by comparing it with the calculations obtained from the IBM Quantum device. Fig. \ref{fig:rf}b shows the population distribution as a function of $\theta_{y}$. Initially (at $\theta_{y}=0$) all the population exists only in $m_{f}=0$ state. As we increase $\theta_{y}$ the population transfer initiates. At $\theta_{y}=\frac{\pi}{2}$ we see half of the population is in $m_{f}=0$ and the remaining half is equally distributed in $m_{f}=\pm1$. At $\theta_{y} = \pi$ the entire population is distributed equally in $m_{f}=\pm1$ states. After this points if we increase $\theta_{y}$ the $m_{f}=0$ population increases again and shows a symmetric behaviour as expected. The population distribution shown in Fig. \ref{fig:rf}b goes well with the experimentally observed results in Fig 1.(D) of \cite{kondakci2020interferometric} for the time scale of 40 $\mu s$.

 As a result of RF coupling which results in population transfer to $m_f=1$ and $m_f=-1$ from initially created $m_f=0$ state the spin part of the scattered wavefunction for a single BEC eventually ends up in a superposition described by: 

 \begin{equation}
    \ket{\phi_a}=C^{'}_{-1}\ket{1,-1}_{a}+C^{'}_0\ket{1,0}_{a}+C^{'}_1\ket{1,1}_{a}
    \label{rfeqn1}
\end{equation}

Where $C^{'}_{-1}$, $C^{'}_{0}$ and $C^{'}_{1}$ are the coefficients of superposition, the first and the second number inside the Ket denotes F=1 and the associated hyperfine spin ($m_{f}$) respectively. Eq.\ref{rfeqn1} shows that the superposition created via RF does not have any momentum parts and thus is much simpler to work with. The reaction rates ratios for both the reaction channels does not change and Eq.\ref{des} and Eq.\ref{cons} still apply.  

\begin{figure}[h]
\includegraphics[width=1.\linewidth]{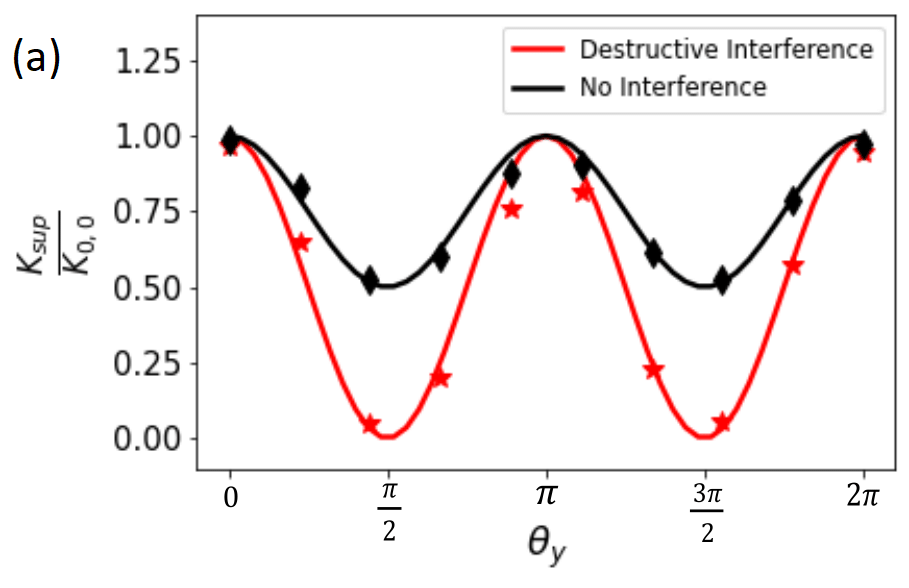}
 \quad
  \includegraphics[width=1.\linewidth]{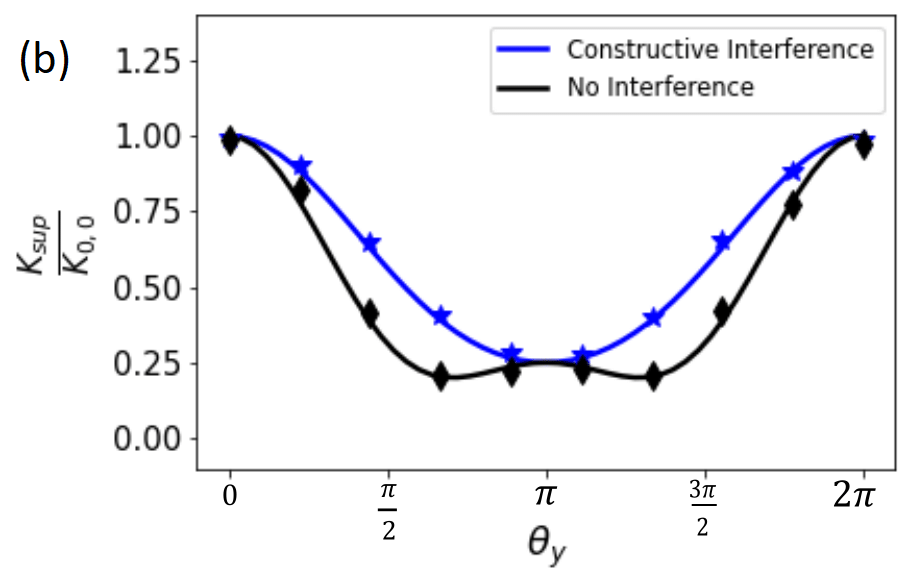}
 \caption{\label{fig:rf_rates}%
  Photo-association rates ratio $k_{sup}/k_{0,0}$ of BEC, as a function of $\theta_{y}$, fig.\ref{fig:rf_rates}a shows the curve for the channel $\ket{F=0,m_{f}=0}$. The black(red) curve corresponds to theoretical prediction without(with) the destructive interference term in the Eq.\ref{des} and the black diamond(red asterisk) corresponds to data points obtained from the IBM Device. The curve for the channel $\ket{F=2,m_{f}=0}$ is shown in fig.\ref{fig:rf_rates}b. The black(blue) curve corresponds to theoretical prediction without(with) the constructive interference term in the Eq. \ref{cons} and the black diamond(blue asterisk) corresponds to data points obtained from the IBM quantum device (IBMQ Lima).
 }%
\end{figure}

Fig.\ref{fig:rf_rates}a shows the normalized Photo-association rates ratio $k_{sup}/k_{0,0}$ of BEC for the $\ket{F=0,m_{f}=0}$ channel, as a function of rotation along Y axis ($\theta_{y}$) which ranges from 0 to $2\pi$. At $\theta_{y}=0$ all the population exists in $m_{f}=0$ and thus we don't have superposition in the reactant state, which corresponds to $C^{'}_{0}=1$ and $C^{'}_{\pm1}=0$ in Eq.\ref{rfeqn1} and Eq.\ref{des}. Thus $k_{sup}/k_{0,0}\longrightarrow 1$. As we increase the angle of rotation $\theta_{y}$ the population transfer takes place as shown in fig.\ref{fig:rf}b and the reaction rate ratio drastically falls down for the case in which we considered the interference (red curve/asterisks). At the rotation angle $\theta_{y}=\pi/2$ the reaction is completely suppressed for the interference case. The reason being at this point half of the population is present in $m_{f}=0$ state and another half is equally distributed in $m_{f}=\pm1$ states, which corresponds to $C^{'}_{0}=1/\sqrt{2}$ and $C^{'}_{\pm1}=1/2$ in Eq.\ref{des} leading to $k_{sup}/k_{0,0}\longrightarrow 0$. At $\theta_{y}=\pi$ the reaction rate ratio for the interference case $k_{sup}/k_{0,0}\longrightarrow 1$ but the reason is entirely opposite to that of $\theta_{y}=0$ case, at this point all the population is equally distributed in $m_{f}=\pm1$ states which corresponds to $C^{'}_{0}=0$ and $C^{'}_{\pm1}=1/\sqrt{2}$. After this point the trend repeats as expected from the population distribution. In general the reaction rate ratio for the case where we consider interference (red curve) is always less (or equal as explained) compared to the case in which we don't consider interference term (black cure) in Eq.\ref{des}, this is interpreted as destructive interference.  

Similarly Fig.\ref{fig:rf_rates}b shows the normalized PA rates ratio $k_{sup}/k_{0,0}$ of BEC for the $\ket{F=2,m_{f}=0}$ channel, as a function of rotation along Y axis ($\theta_{y}$). At $\theta_{y}=0$, $C^{'}_{0}=1$ and $C^{'}_{\pm1}=0$ and this corresponds to $k_{sup}/k_{0,0}\longrightarrow 1$ from Eq.\ref{cons}. As we increase the angle of rotation $\theta_{y}$ the population transfer takes place. At $\theta_{y}=\pi$ the reaction rate ratio for the case where we consider interference (blue curve) matches with the case where we do not consider interference. The reason being at this point all the population is equally distributed in $m_{f}=\pm1$ which corresponds to $C^{'}_{0}=0$ and $C^{'}_{\pm1}=1/\sqrt{2}$. So out of two reaction pathways only one of them exists ($m_{f}=\pm1$ + $m_{f}=\mp1 \longrightarrow m_{tot} = 0$). It is important to note that Fig.\ref{fig:rf_rates}a and Fig.\ref{fig:rf_rates}b are symmetric about $\theta_{y}=\pi$ which comes from the nature of population distribution in Fig.\ref{fig:rf}b. Also the periodicity of reaction rates in Fig.\ref{fig:rf_rates}a and Fig.\ref{fig:rf_rates}b is $\pi$ and $2\pi$ respectively. The origin of these periodicities is purely numerical and depends on the rate expressions Eq.\ref{des}, Eq.\ref{cons} and the population distribution Fig\ref{fig:rf}b. In general the reaction rate ratio for the case where we consider interference (blue curve) is always greater (or equal as explained) compared to the case in which we don't consider interference term (black cure) in Eq.\ref{cons}. This is interpreted as constructive interference.

In summery there are multiple approaches to achieve constructive interference within PA reaction. For example the recent study by Esat et. al. \cite{kondakci2020interferometric} showed the interferometric control over the $\ket{F=0,m_{f}=0}$ reaction channel by exploiting the quadratic zeeman shift which introduces an additional relative phase between $m_{f}=0$ and $m_{f}=\pm{1}$ hyperfine spins in the superposition state Eq.\ref{eq2}. We have showed that by changing the scattering channel from $\ket{F=0,m_{f}=0}$ to $\ket{F=2,m_{f}=0}$ we can achieve a constructive interference. The reason behind this result is similar(opposite) sign of CG coefficients in the latter(former). We are investigating the existence of a spin sensitive PA frequency \cite{hamley2009photoassociation} which corresponds to the  $\ket{F=2,m_{f}=0}$ scattering channel. Our study shows that quantum interferences can be employed to coherently control a photo-chemical reaction. The approach is general and can be used to study a wide range of chemical reactions in the ultra-cold regime. Next we plan to investigate the role of entanglement \cite{karra2016prospects,li2019entanglement,kais2007entanglement} to control and predict the interference patterns observed in different scattering experiments which are similar to the PA reaction of SOC BEC\cite{liu2021precision,liu2020precision,sneha2016multiple}.

% The \nocite command causes all entries in a bibliography to be printed out
% whether or not they are actually referenced in the text. This is appropriate
% for the sample file to show the different styles of references, but authors
% most likely will not want to use it.
%\nocite{*}
\bibliography{apssamp}% Produces the bibliography via BibTeX.

\widetext
\section*{Supplemental Materials}
Scattered wavefunction for a single particle in the BEC which is loaded into a single particle ground state with quasi momentum $ \Vec{q} = (0,q_{min},0)$ (without loss of generality, we have assumed that the raman beams are applied along Y direction as shown in the Fig. 1(b) of Ref.\cite{blasing2018observation}) is   
 \begin{equation}
    |\psi_a\rangle=C_{-1} e^{i(\Vec{q}-\Vec{k}_{r}).\Vec{r}_{a}}\ket{1,-1}_{a}+C_{0} e^{i(\Vec{q}).\Vec{r}_{a}}\ket{1,0}_{a}+ C_{1} e^{i(\Vec{q}+\Vec{k}_{r}).\Vec{r}_{a}}\ket{1,1}_{a}
    \label{si_eq2}
\end{equation}
Here kets with subscript $a$ and $r_{a}$, denotes the spin state and the coordinate of the single atom respectively and $\Vec{k_r}=(0,-2k_{r},0)$. The product state of two such particles ( here denoted as $a$ and $b$) which involves superposition in the spin portion of the scattering wavefunction  (with some extra spatial dependent phases due to Raman coupling), can be expressed as:

\begin{equation}
         e^{i\Vec{q}.(\Vec{r}_{a}+\Vec{r}_{b})} (C_{0}^{2} \ket{1,0}_{a}\ket{1,0}_{b} + C_{1}C_{-1} e^{\Vec{k}_{r}.(\Vec{r}_{a}-\Vec{r}_{b})} \ket{1,1}_{a}\ket{1,-1}_{b} +C_{-1}C_{1} e^{i\Vec{k}_{r}.(\Vec{r}_{b}-\Vec{r}_{a})} \ket{1,-1}_{a}\ket{1,1}_{b}+ ...)
\end{equation}

After using proper Clebesch-Gordon coefficients \cite{griffiths:quantum} the product state wavefunction can be written as: 
\begin{flalign}
    &= e^{i\Vec{q}.(\Vec{r}_{a}+\Vec{r}_{b})} [C_{0}^{2} (\sqrt{\frac{2}{3}} \ket{2,0} -\sqrt{\frac{1}{3}} \ket{0,0} )&&\\\nonumber
            & \hspace{0.5cm} + C_{1}C_{-1} e^{i\Vec{k}_{r}.(\Vec{r}_{a}-\Vec{r}_{b})} (\sqrt{\frac{1}{6}} \ket{2,0}-\sqrt{\frac{1}{2}} \ket{1,0}+\sqrt{\frac{1}{3}} \ket{0,0})&&\\\nonumber
            & \hspace{0.5cm} + C_{-1}C_{1} e^{i\Vec{k}_{r}.(\Vec{r}_{b}-\Vec{r}_{a})} (\sqrt{\frac{1}{6}} \ket{2,0}-\sqrt{\frac{1}{2}} \ket{1,0}+\sqrt{\frac{1}{3}} \ket{0,0})
    +...]
    &&\\\nonumber
    &= e^{i\Vec{q}.(\Vec{r}_{a}+\Vec{r}_{b})} \left[\sqrt{\frac{2}{3}}C_{0}^{2}+\sqrt{\frac{1}{6}}C_{1}C_{-1}e^{i\Vec{k}_{r}.(\Vec{r}_{a}-\Vec{r}_{b}))}+\sqrt{\frac{1}{6}}C_{-1}C_{1}e^{i\Vec{k_{r}}.(\Vec{r}_{b}-\Vec{r}_{a}))}\right] \ket{2,0} + ... &&\\\nonumber
    &= e^{i\Vec{q}.(\Vec{r}_{a}+\Vec{r}_{b})} \left[\sqrt{\frac{2}{3}}C_{0}^{2}+\sqrt{\frac{1}{6}}C_{1}C_{-1}e^{i\Vec{k}_{r}.\Vec{r}_{ab}}+\sqrt{\frac{1}{6}}C_{-1}C_{1}e^{-i\Vec{k}_{r}.\Vec{r}_{ab}}\right] \ket{2,0} + ... &&\\\nonumber    
\end{flalign}

Here kets with subscript $a$ or $b$, denotes the spin states of the two single atoms respectively, and the ones without subscripts correspond to the total spins of two particles. $\Vec{r}_{ab}$ is the relative coordinate which stands for $\Vec{r}_{a}-\Vec{r}_{b}$.

For the dressed atoms in the scattering channel $\ket{F=2,m_{F}=0}$, the stimulated transition rate to the excited molecular state is:
\begin{equation}
    \Gamma_{sup} \propto |\braket{\phi_{m}(\vec{r}_{ab})}{\bra{F=2,m_{f}=0}\psi_{scat}}|^{2}
    \label{rate}
\end{equation}

Here $\psi_{scat}$ refers to the two body scattering wavefunction of the colliding atoms, which includes both spatial and spin parts. The operator $\bra{F=2,m_{f}=0}$ selects only that part of $\psi_{scat}$ which corresponds to the chosen photoassociation channel with total spin $\ket{F=2,m_{f}=0}$. $\phi_{m}(\vec{r}_{ab})$ and $\phi_{F=2}(\vec{r}_{ab})$ corresponds to the spatial wavefunctions for the molecules and the bare scattering state in the allowed Channel. The total scattering wavefunction becomes:
\begin{equation}
    \ket{\psi_{scat}} =  \left[\sqrt{\frac{2}{3}}C_{0}^{2}+\sqrt{\frac{1}{6}}C_{1}C_{-1}e^{i\Vec{k}_{r}.\Vec{r}_{ab}}+\sqrt{\frac{1}{6}}C_{-1}C_{1}e^{-i\Vec{k}_{r}.\Vec{r}_{ab}}\right] \ket{2,0} \phi_{F=2}(\vec{r}_{ab}) + ...
\end{equation}

We have suppressed the $e^{i\Vec{q}.(\Vec{r}_{a}+\Vec{r}_{b})}$ as it's an overall phase. In addition we multiplied the spacial wave function $\phi_{F=2}(\vec{r}_{ab})$ with the corresponding contributing spin product wavefunction. ... are the remaining projections of the scattered wavefunction which does not contribute to the selected channel. Now we project $\ket{\psi_{scat}}$ to include only the portion with $\ket{F=2,m_{F}=0}$:

\begin{equation}
    \bra{F=2,m_{F}=0}\ket{\psi_{scat}} = \left[\sqrt{\frac{2}{3}}C_{0}^{2}+\sqrt{\frac{1}{6}}C_{1}C_{-1}e^{i\Vec{k}_{r}.\Vec{r}_{ab}}+\sqrt{\frac{1}{6}}C_{-1}C_{1}e^{-i\Vec{k}_{r}.\Vec{r}_{ab}}\right] \phi_{F=2}(\vec{r}_{ab})
\end{equation}

The stimulated transition rate Eq. \ref{rate} becomes,

\begin{equation}
\Gamma_{sup} \propto \left| \frac{\sqrt{2}C_{0}^{2}}{\sqrt{3}}  \left( \int f(\vec{r}_{ab})\, d\vec{r}_{ab} \right) + \frac{C_{1}C_{-1}}{\sqrt{6}} \left( \int f(\vec{r}_{ab})\,e^{i\Vec{k}_{r}.\Vec{r}_{ab}}\, d\vec{r}_{ab} + \int f(\vec{r}_{ab})\,e^{-i\Vec{k}_{r}.\Vec{r}_{ab}}\, d\vec{r}_{ab} \right) \right|^{2}
\label{big_half}
\end{equation}

Where the function $f(\vec{r}_{ab})$ is $\phi_{m}^{*}(\vec{r}_{ab})\, \phi_{F=2}(\vec{r}_{ab})$, the integral in Eq.\ref{big_half} contains the Frank-Condon overlap and the additional phases associated with Raman beams weighted by the corresponding superposition coefficients. We consider $\vec{k}_{r}.\vec{r}_{ab}$ negligible and it's justifiable since the size of our molecule is $<< \lambda_{R}$ (where $\lambda_{R} = \frac{2\pi}{k_{r}} \approx 15000\,a_{o}$). Franck-Condon overlap integrals depends only by the short-range behavior, then 

\begin{equation}
\int \phi_{m}^{*}(\vec{r}_{ab})\, \phi_{F=2}(\vec{r}_{ab})\,e^{i\Vec{k}_{r}.\Vec{r}_{ab}}\, d\vec{r}_{ab} \approx \int \phi_{m}^{*}(\vec{r}_{ab})\, \phi_{F=2}(\vec{r}_{ab})\, d\vec{r}_{ab} \approx \int \phi_{m}^{*}(\vec{r}_{ab})\, \phi_{F=2}(\vec{r}_{ab})\,e^{-i\Vec{k}_{r}.\Vec{r}_{ab}}\, d\vec{r}_{ab}
\label{approx}
\end{equation}

Therefore the stimulated transition rate in Eq. \ref{big_half} reduces to
\begin{equation}
\Gamma_{sup} \propto \frac{2}{3} \left|\int \phi_{m}^{*}(\vec{r}_{ab})\, \phi_{F=2}(\vec{r}_{ab})\, d\vec{r}_{ab}  \right|^{2} \left| C_{0}^{2}+C_{1}C_{-1}\right|^{2}
\end{equation}

For the case when we don't have superimposed reactant state (When $C_{0}=1$ and $C_{\pm1}=0$) the particles are in spin sate $\ket{f=0,m_{f}=0}$ for this case the stimulated rate of two particles in the scattering channel $\ket{F=2,m_{F}=0}$ is denoted as $\Gamma_{0,0}$, in this case they have projection along  $\ket{F=2,m_{F}=0}$ with CG coefficient of $\sqrt{\frac{2}{3}}$, then:
\begin{equation}
\Gamma_{0,0} \propto \frac{2}{3}\left| \int \phi_{m}^{*}(\vec{r}_{ab})\, \phi_{F=2}(\vec{r}_{ab})\, d\vec{r}_{ab}\right|^{2} 
\end{equation}

Therefore,  $\Gamma_{sup}/\Gamma_{0,0}=\left| C_{0}^{2}+C_{1}C_{-1}\right|^{2}$. We know that $k_{sup} \propto \Gamma_{sup}$ with a proportionality constant which is independent of the spin of colliding atoms \cite{theis2004tuning}. 

\begin{equation}
\frac{k_{sup}}{k_{0,0}} = \left| C_{0}^{2}+C_{1}C_{-1}\right|^{2} = \left|C_{0}^{2} \right|^{2}+\left|C_{1}C_{-1} \right|^{2} + 2\,Re[C_{0}^{2}C_{1}^{*}C_{-1}^{*}]
\label{cons}
\end{equation}

\end{document}